# EMERGENCE, COMPETITION AND DYNAMICAL STABILIZATION OF DISSIPATIVE ROTATING SPIRAL WAVES IN AN EXCITABLE MEDIUM: A COMPUTATIONAL MODEL BASED ON CELLULAR AUTOMATA

S. D. Makovetskiy, D. N. Makovetskii*

*Institute for Radiophysics and Electronics of the National Academy of Sciences of Ukraine, Kharkov, Ukraine

Electronic addresses:
sdmakovetskiy@mail.ru
makov@ire.kharkov.ua

**Abstract.** In the present survey paper, we report some qualitatively new features of emergence, competition and dynamical stabilization of dissipative rotating spiral waves (RSWs) in the cellular-automaton model of laser-like excitable media proposed in arXiv:cond-mat/0410460v2, arXiv:cond-mat/0602345. Part of the observed features are caused by unusual mechanism of excitation vorticity when the RSW's core gets into the surface layer of an active medium. Instead of the well known scenario of RSW collapse, which takes place after collision of RSW's core with absorbing boundary, we observed complicated transformations of the core leading to regeneration (nonlinear "reflection" from the boundary) of the RSW or even to birth of several new RSWs in the surface layer. Computer experiments on bottlenecked evolution of such the RSW-ensembles (vortex matter) are reported and a possible explanation of real experiments on spin-lattice relaxation in dilute paramagnets is proposed on the basis of an analysis of the RSWs dynamics. Chimera states in RSW-ensembles are revealed and compared with analogous states in ensembles of nonlocally coupled oscillators. Generally, our computer experiments have shown that vortex matter states in laser-like excitable media have some important features of aggregate states of the usual matter.

**Introduction**

Dissipative rotating spiral waves (RSWs) are robust dissipative structures observed not only in diverse fields of the fundamental science, but in several types of man-made active media (or arrays of units) used in quantum electronics, nanoelectronics etc. All these RSW are vortex states, so they have the certain topological charges similar to charges of "the singular optics" vortices. But, in contrast to *conservative* singular optics vortices, *dissipative* RSW are self-sustained spatio-temporal structures, called also autowaves. In particular, they possess a responsibility of non-decaying propagation in active media (maintained by the balance of an appropriate pumping and corresponding dissipation of energy).

Evolution of various spatio-temporal structures of the RSW type is now a subject of extensive experimental investigations, theoretical studies and computer modeling (see, e.g., [1], [2], [3], [4], [5], [6], [7] and references therein). A wide class of mathematical models in this field is based on the concept of excitability [8].

An important approach in modern computer investigations of excitable systems is based on using of discrete parallel models with local interactions, mostly by cellular automata (CA). In the framework of the CA approach [9], the three-level representation of the excitable media states is an effective tool for carrying computer experiments with multi-particle systems possessing long-time nonstationary dynamics (see [6], [7] and references therein). Quantity of particles (elements) $D$ may be mesoscopically large: $D = 10^3 - 10^6$ and more.

Parallelization of updating of the active medium states, locality of the interactions between elementary parts of the medium and the full discreteness (space-time-state) of the system are the fundamental principles of the CA approach [9]. Using of this approach admits to fulfill modeling of large systems by emulation of parallel matrix transformations. The main goal of this work is a CA modeling of dynamics of ensembles of the dissipative RSW in bounded excitable media with three-level discrete active units. A near real-world example of such the system is the phaser (microwave phonon laser) [10], [11], [12], [13], elaborated in Institute of Radio-Physics and Electronics (Kharkov, Ukraine) on the basis of theoretical ideas and experimental methods proposed in early 1960th [14], [15], [16], [17], [18]. For recent experimental data on the phaser dynamics, see [19], [20], [21], [22], [23], [24], [25].

**1. Three-level model of an excitable system – the general approach**

The essence of the three-level representation of a parametrically homogeneous and isotropic (in the von Neumann sense [26]) excitable medium is as follows [27], [28], [29]. Each active element (AE) of an excitable system has the single stable *ground* state (low-lying level), say $L_{\text{I}}$, and two metastable states – upper levels, say $L_{\text{II}}$ and $L_{\text{III}}$. The highest metastable state $L_{\text{III}}$ is the *excited* one, it reaches only by certain hard perturbation of the ground state $L_{\text{I}}$ of the considered AE. Such perturbation may be caused both by an external source or by interaction with another AEs, which belong to the active vicinity (AV) of the considered AE.

The excitation of the state $L_{\text{III}}$ (i.e. the transition $L_{\text{I}} \to L_{\text{III}}$) is a kind of purely induced transition. Due to metastability, the decay time $\tau_d$ of the excited state $L_{\text{III}}$ is a finite value $\tau_d \leq \tau_e$, after which every excited AE reaches the intermediate *refractory* state $L_{\text{II}}$. Here, $\tau_e$ is the lifetime of an *isolated* AE in the $L_{\text{III}}$ state.

For the simplest case, the transition $L_{\text{III}} \to L_{\text{II}}$ is a purely spontaneous one [29], so $\tau_d = \tau_e$. Generally, the state $L_{\text{III}}$ of an AE may loose its stability at $\tau = \tau_d < \tau_e$ (or even at $\tau = \tau_d \ll \tau_e$) due to interaction with surrounding (located in the AV) ground-state ACs [6], [7]. In the last case, the decay time $\tau_d$ is defined by current states of all the ACs belonging to the AV.

Refractority means "sleeping" state at which no excitations of the considered AE are possible both from neighbouring ACs and from an external source. Moreover, such the AE (lying at $L_{\text{II}}$) can't participate in excitation of any another AE (lying at $L_{\text{I}}$). After a refractority lifetime $\tau_r$, the considered AE reaches the ground state $L_{\text{I}}$ without reference to states of the neighbouring AE. In other words, the transition $L_{\text{II}} \to L_{\text{I}}$ is the purely spontaneous one. Only the cyclic transitions $L_{\text{I}} \to L_{\text{III}} \to L_{\text{II}} \to L_{\text{I}} \to \ldots$ are permitted in the framework of the canonical three-level model [27], [29] of excitable system.

**2. Laser-like analogs of excitable systems**

The concept of excitability was primarily developed in biology, but it is widely used now in chemistry and physics. In particular, various realizations of this concept have been applied to solve problems arising in modern nonlinear optics, laser physics etc. [30], [31], [32], [33].

In [6], [7], [34], [35], [36], [37], we have considered even more close laser-like analog of excitable system, namely the three-level active medium of the microwave phonon laser (phaser) with dipole-dipole interactions between AEs. The only significant difference of the phaser medium from the usual excitable one is the possibility of the two-channel (generally, multi-channel) diffusion of excitations. This system was experimentally and theoretically studied by D. N. Makovetskii [12], [13], [19], [20], [21], [22], [23], [24], [25], demonstrating self-organization, bottlenecked cooperative transient processes, coexistence of regular and irregular spatio-temporal dissipative structures and other nonlinear phenomena under extremely low level of intrinsic quantum noise (phaser has 15 orders lower intensity of spontaneous emission in comparison to usual optical-range lasers).

In preceding publications [6], [7], [34], [35], [36], we have reported on a series of computer experiments concerning the RSW dynamics in autonomous excitable systems with various control parameters and random initial excitations. The most interesting phenomena of self-organized vorticity observed in these computer experiments were as follows:

(a) spatio-temporal transient chaos in the form of highly bottlenecked collective evolution of the vortex matter formed by RSWs with variable topological charges;

(b) competition of left-handed and right-handed RSWs with unexpected features, including self-induced inversion of the topological charges ("right-glove $\leftrightarrow$ left-glove" transformation of spatio-temporal structures);

(c) coexistence of regular (RSW-type) and chaotic (labyrinthine-type) domains in excitable media;

(d) full restoring of spatial symmetry of dissipative structures in an excitable medium after its initialization by a starting pattern with initially broken symmetry (the counterpart of the well-known phenomenon of broken symmetry).

Phenomena (a) and (c) are directly related to microwave phonon laser dynamics features observed earlier in real experiments at liquid helium temperatures on corundum crystals doped by iron-group ions [12], [13], [19], [20], [21], [22], [23], [24], [25].

In the present paper, we report some qualitatively new features of emergence, competition and dynamical stabilization of RSWs in cellular-automaton model of laser-like excitable media [6], [7], [34], [35], [36]. Part of these features are caused by unusual mechanism of RSW evolution when RSW's cores get into the surface layer of an active medium (i. e. the layer of ACs resided at the absorbing boundary). Instead of well known scenario of RSW collapse, which takes place after collision of RSW's core with absorbing boundary, we observed complicated transformations of the core leading to "reflection" of the RSW from the boundary or even to birth of several new RSW in the surface layer. To our knowledge, such nonlinear "reflections" of RSW and resulting die hard vorticity in excitable media with absorbing boundaries were unknown earlier. Computer experiments on bottlenecked evolution of RSW are reported and a possible explanation of real experiments on spin-lattice relaxation in dilute paramagnets is proposed. Chimera states in RSW-ensembles are revealed and compared with analogous states in ensembles of nonlocally coupled oscillators. Generally, our computer experiments have shown that collective vortex matter states in laser-like excitable media have some important features of aggregate states of the usual matter.

### 3. Tools for modeling of dynamics of three-level excitable systems

Most of the discussed below numerical experiments were fulfilled by using the cross-platform software package "Three-Level Model of excitable system" (TLM © 2006 S. D. Makovetskiy) [7], [35], [36] and it's previous (console) variant "Three-Level Cellular Automaton" (TLCA © 2004 S. D. Makovetskiy) [34]. The TLM package is a Java-based implementation [35], [36] of the paramagnetic-oriented modification [6] of the well-known Zykov-Mikhailov (ZM) model [29]. The underlying algorithm of TLM (ATLM) [7], [35], [36] is based on binary/integer/rational numbers and Boolean/arithmetic operations only, so it allows direct mapping of the model onto an architecture of a standard digital computer. The ATLM belongs to the polynomial-time ($P$-time) class of algorithmic complexity; moreover, it is a *Linear*-time algorithm for the case of predefined quantity of iterations $N$ of a modeling session, so it provides possibility of modeling of long evolution of large systems ($ND \approx 10^{12}$ at usual personal computers and $ND \approx 10^{13} - 10^{15}$ at mid-power machines) during a workday. Moreover, the ATLM is intrinsically fine-grained parallel algorithm, based on CA approach, which resolves in a natural way the contradiction between efficiency and parallelization in computational mathematics. This property of CA gives possibility of further increasing of the parameter $ND$ [36], e.g. by using clusters of computers or by using specialized co-processors with fine-grained hardware architecture and with built-in task-oriented firmware.

The ATLM [7], [35], [36] and the corresponding TLM software package [35], [36] explicitly include a mechanism of spatially-dependent local inhibition of AE excitations in diluted paramagnetic excitable system [6], [7]. The original ZM model [29] takes into account only the usual spatially-independent mechanism of inhibition (which is an analogue of spin-lattice relaxation in paramagnetic systems), so the chemical-oriented ZM model of excitable systems [29] disregards any cross-relaxation phenomena which are typical for paramagnetic excitable systems. It must be pointed out that not only ZM model, but almost all models of Wiener-Rosenblueth type [29] ignore spatial dependence of inhibition, restricting themselves by spatial dependence of local activation only. This approach is usually correct for chemical systems (e.g. of Belousov-Zhabotinsky type), but it is inadmissible for paramagnetic excitable systems. In works [6], [36], it was shown that taking into account both local activation and local inhibition of excitation is equal to generalization of the CA model with the 1-channel (1C) diffusion to the 2-channel (2C) one. Generalization of this approach gives possibility to model excitable-like systems with more than three levels and more than two channels of diffusion.

### 4. Explicit formulation of the algorithm used in our computer experiments

The generic algorithm for 1C-diffusion in discrete excitable media with three-state AEs and non-zero times of relaxation was firstly proposed by V.S. Zykov and A.S. Mikhailov (ZM) in [29]. A modification of ZM algorithm, which allows modeling of 2C-diffusion in excitable media (and which may be used for investigation of spatio-temporal structures in class-B lasers), was for the first time proposed in a series of publications [6], [7], [34], [35], [36]. The essence of this algorithm (called here ATLM) is as follows [6], [7]:

Let an active (excitable) discretized medium $P_E$ has the form of the rectangular 2D-lattice. Each cell of the lattice contains one AE with coordinates $i, j$. All the AEs in the $P_E$ are parametrically identical. The upgrade of state $S_{ij}^{(n)}$ of each AE is carried out synchronously during the evolution. This upgrade is defined by rules depending on a level $L_K$ at which the AE is located at the step $n$. Every AE at each step of evolution $n$ is at one of three discrete levels $L_K \in \{L_I; L_{II}; L_{III}\}$. The complete description of AE's state $S_{ij}^{(n)}$ in our three-level model (TLM) of excitable system includes one global attribute (the phase counter $\varphi_{ij}^{(n)}$) and two partial attributes $u_{ij}^{(n)}$ and $z_{ij}^{(n)}$ for each individual AE in $P_E$. Full description of all AE's possible states is as follows:

$$S_{ij}^{(n)}(L_I) = (\varphi_{ij}^{(n)}, u_{ij}^{(n)}); \quad S_{ij}^{(n)}(L_{III}) = \begin{cases} (\varphi_{ij}^{(n)}), & \text{if } n = 0; \\ (\varphi_{ij}^{(n)}; z_{ij}^{(n)}), & \text{if } n \neq 0. \end{cases}; \quad S_{ij}^{(n)}(L_{II}) = (\varphi_{ij}^{(n)}). \quad (1)$$

In the framework of the TLM model, the phase counters $\varphi_{ij}^{(n)} \in [0, \tau_e + \tau_r]$ and the following correspondences between $\varphi_{ij}^{(n)}$ and $L_K$ take place for all the AEs in $P_E$ at all steps of evolution:

$$(L_K = L_I) \Leftrightarrow (\varphi_{ij}^{(n)} = 0); \quad (L_K = L_{III}) \Leftrightarrow (0 < \varphi_{ij}^{(n)} \leq \tau_e); \quad (L_K = L_{II}) \Leftrightarrow (\tau_e < \varphi_{ij}^{(n)} \leq \tau_e + \tau_r). \quad (2)$$

The evolution of the individual AE proceeds by transitions $L_K \to L_{K'}$ (where $K, K' \in \{I; II; III\}$) induced by the Kolmogorov operator $\hat{\Omega}$. In TLM model, the $\hat{\Omega}$ operator has three orthogonal branches $\hat{\Omega}_I$, $\hat{\Omega}_{III}$ and $\hat{\Omega}_{II}$, which change or not the state (i.e. the level $L_I$, $L_{III}$ or $L_{II}$) of each AE (depending on $\varphi_{ij}^{(n)}$):

$$\varphi_{ij}^{(n+1)} = \hat{\Omega}\varphi_{ij}^{(n)} = \begin{cases} \hat{\Omega}_I \varphi_{ij}^{(n)}, & \text{if } \varphi_{ij}^{(n)} = 0; \\ \hat{\Omega}_{III} \varphi_{ij}^{(n)}, & \text{if } 0 < \varphi_{ij}^{(n)} \leq \tau_e; \\ \hat{\Omega}_{II} \varphi_{ij}^{(n)}, & \text{if } \tau_e < \varphi_{ij}^{(n)} \leq \tau_e + \tau_r, \end{cases} \quad (3)$$

where the lifetimes for the excited and refractory levels $\tau_e$ and $\tau_r$ are integer positive numbers (note that in the original WR-model [27] $\tau_e = 0$; a modified WR-model with $\tau_e \neq 0$ was proposed later by A. Rosenblueth [28]).

So, at step $n+1$, the branch $\hat{\Omega}_I$ does operations only with those AEs, which have $L_K = L_I$ at step $n$:

$$\varphi_{ij}^{(n+1)} = \begin{cases} 0, \text{if } (\varphi_{ij}^{(n)} = 0) \wedge (u_{ij}^{(n+1)} < h); \\ 1, \text{if } (\varphi_{ij}^{(n)} = 0) \wedge (u_{ij}^{(n+1)} \geq h). \end{cases} \quad \text{with} \quad u_{ij}^{(n+1)} = g u_{ij}^{(n)} + \sum_{p,q} C(p,q) J_{i+p, j+q}^{(n)}, \quad (4)$$

where $h$ is the threshold for the $u$-agent ($h > 0$); $C(p,q)$ is the active neighbourhood of the AE at $(i, j)$; $J_{i+p, j+q}^{(n)} \in \{0; 1\}$; $g \in [0, 1]$; and we assume that $u$-agent arrives from AEs with $L_K = L_{III}$ in $C(p,q)$:

$$J_{i+p, j+q}^{(n)} = \begin{cases} 1, \text{if } 0 < \varphi_{i+p, j+q}^{(n)} \leq \tau_e; \\ 0, \text{if } (\tau_e < \varphi_{i+p, j+q}^{(n)} \leq \tau_e + \tau_r) \vee (\varphi_{i+p, j+q}^{(n)} = 0). \end{cases} \quad (5)$$

In this work, we use the active neighbourhoods of the Moore ($C = C_M$) and the von Neumann ($C = C_N$) types:

$$C_M(p,q) = \begin{cases} 1, \text{ if } [(|p| \leq 1) \wedge (|q| \leq 1) \wedge (\delta_{p0}\delta_{q0} \neq 1)]; \\ 0, \text{ otherwise}. \end{cases} ; \quad C_N(p,q) = \begin{cases} 1, \text{ if } [(|p|=1) \oplus (|q|=1)]; \\ 0, \text{ otherwise}. \end{cases}, \quad (6)$$

where $\delta_{ij}$ is the Kronecker symbol, and the symbol $\oplus$ means XOR (excluding OR).

At the same step $n+1$, the branch $\hat{\Omega}_{\text{III}}$ does operations only with AEs having $L_K = L_{\text{III}}$ at step $n$:

$$\varphi_{ij}^{(n+1)} = \begin{cases} \varphi_{ij}^{(n)} + 1, \text{ if } [(0 < \varphi_{ij}^{(n)} < \tau_e) \wedge (z_{ij}^{(n+1)} < f)] \vee (\varphi_{ij}^{(n)} = \tau_e); \\ \varphi_{ij}^{(n)} + 2, \text{ if } (0 < \varphi_{ij}^{(n)} < \tau_e) \wedge (z_{ij}^{(n+1)} \geq f). \end{cases} \text{ with } z_{ij}^{(n+1)} = \sum_{p,q} C(p,q) Q_{i+p,j+q}^{(n)}, \quad (7)$$

where $f$ is the threshold for the $z$-agent ($f > 0$); $Q_{i+p,j+q}^{(n)} \in \{0;1\}$, and we assume that $z$-agent arrives (to AE at $(i,j)$ with $L_K = L_{\text{III}}$) from AEs with $L_K = L_{\text{I}}$ in $C(p,q)$:

$$Q_{i+p,j+q}^{(n)} = \begin{cases} 1, \text{ if } \varphi_{i+p,j+q}^{(n)} = 0; \\ 0, \text{ if } \varphi_{i+p,j+q}^{(n)} \neq 0. \end{cases} \quad (8)$$

One can see from (7) that the $z$-agent does not accumulate during successive iterations. In other words, the branch $\hat{\Omega}_{\text{III}}$ at step $n+1$ produces new values of partial attributes $z_{ij}^{(n+1)}$ (for AEs having $L_K = L_{\text{III}}$ at step $n$ — in contrary to the $u$-agent for AEs having $L_K = L_{\text{I}}$ at step $n$).

Finally, the branch $\hat{\Omega}_{\text{II}}$ does not produce/change any partial attributes at all (because the intermediate level $L_{\text{II}}$ always is in the state of refractority). It does such the operations with AEs having $L_K = L_{\text{II}}$ at step $n$:

$$\varphi_{ij}^{(n+1)} = \begin{cases} \varphi_{ij}^{(n)} + 1, \text{ if } \tau_e < \varphi_{ij}^{(n)} < \tau_e + \tau_r; \\ 0, \quad \text{ if } \varphi_{ij}^{(n)} = \tau_e + \tau_r. \end{cases} \quad (9)$$

The described algorithm (ATLM) includes 2C-transitions for cases of the type ($C = C_M$, $f \leq 8$) or ($C = C_N$, $f \leq 4$). It may be used for, particularly, modeling of spatio-temporal dynamics in class-B phasers with dipole-dipole interactions between active centers. At ($C = C_M$, $f > 8$), the ATLM becomes equivalent to the ZM-algorithm [29], and at ($C = C_N$, $f > 4$) it is of ZM-like (i.e. 1C) type. In our computer experiments we used $J_{i+p,j+q}^{(n)} \in \{0;1\}$ and $C(p,q) \in \{0;1\}$. For the pointed situation and with $g \in \{0;1\}$, the ZM-algorithm [29] is obviously of fully integer kind. If the last three conditions are supplemented by $Q_{i+p,j+q}^{(n)} \in \{0;1\}$, the ATLM becomes fully integer too. Finally, at ($C = C_M$, $f > 8$, $h = 1$, $g = 0$) the ATLM describes the original Wiener-Rosenblueth model (in the case of its discretized form, which was proposed in [29]).

### 5. Regeneration and replication of dissipative RSW in bounded media

In computer experiments fulfilled by the authors in 2004-2005 [6], there were observed several new scenarios of self-organization, competition and dynamical stabilization of RSW under conditions of cross-relaxation between AEs. Subsequent careful investigations carried out by S. D. Makovetskiy [7], [35], [36] on excitable media with 2C-diffusion have shown that interaction of RSW with absorbing boundaries possesses peculiarities, which were unknown for such media with 1C-diffusion. The most interesting results obtained in [7], [35], [36] – revealing of regeneration and replication of RSW in excitable systems with 2C-diffusion.

Regeneration of RSWs in a surface layer of bounded excitable medium looks as "reflection" of spiral autowave from a boundary. This, of course, is not a trivial reflection because of non-reflective (fully absorbing) nature of the boundary. Moreover, in contrary to ordinary waves, autowaves generally cannot be reflected in usual way (even by perfect mirror). As a matter of fact, nonlinear "reflections" observed in computer experiments [7], [35], [36] were identified as complex *nonlinear transformations* of RSW leading to inversion of direction of drift of a revived and self-reconstructed RSW. In some cases, not only this direction of RSW core motion, but the sign of the RSW's topological charge $Q_T$ was inverted too [7], [35], [36]. This last result correlates with observations of self-organized inversion of $\text{sgn} Q_T$ in preceding computer experiments [6], [34].

Replications of RSWs [7], [35], [36] (increasing of quantity of spirals after their nonlinear transformation in the surface layer) is a higher-order phenomenon of the same nature as regeneration of RSW. Regenerations and multiple replications of spiral autowaves clarify phenomenon of transient Zeno-like spatio-temporal chaos (TZSChaos) which appeared to be typical [7], [36] in investigated excitable media with 2C-diffusion (i.e. with competing local activation and local inhibition of AE excitations). In spite of obvious regularity of all the possi-ble attractors in the TLM model (as in any spatially-bounded cellular automata model), the time of transition process may reach giant values by *almost* precise repetitions of spatial configuration (Zeno-like phenomenon). These repetitions are the consequence of nearly perfect balance between quantity of RSW moving to and from boundaries.

It is important that the computer experiments on regeneration and replication of RSWs [7], [35], [36] were performed in *isotropic and homogeneous* (both in the J. von Neumann sense) media. So the mechanism of self-reconstruction of spiral autowaves in excitable media accompanied by nonlinear "reflection" of RSW and inversion of $\operatorname{sgn} Q_T$ is quite different from superficially similar phenomena of non-autowave optical vorticity in strongly inhomogeneous media [38], [39].

## 6. Bottlenecked evolution of RSW and the puzzle of spin-lattice relaxation

Results of computer experiments on nonstationary phenomena in excitable media emulating paramagnetic systems shed light upon an old but yet non-resolved puzzle of spin-lattice relaxation (SLR) formulated by J.H. Van Vleck in his seminal report at 1st Symposium on Quantum Electronics [40] and widely discussed in many subsequent papers concerning particularly the problem of the nature of bottlenecked SLR in dilute paramagnets at low temperatures. In almost all of these papers, the concept of macroscopic (averaged, thermodynamical etc.) description of bottleneck phenomena in SLR was used. A standard model of bottleneck in SLR is the "phonon bottleneck": SLR is slowed down by lack of resonant phonons, which pass energy from AEs (here, paramagnetic ions) to the thermal reservoir. At the same time, real physical experiments showed inadequacy of this theoretical model. In other words, a description of excited multi-particle paramagnetic system by small set of macroscopic parameters is unsatisfactory.

Taking into account a possibility of spontaneous formation of dissipative structures in paramagnetic media, one can suppose that a bottleneck may originate at the level of particle-particle interaction rather than at the level of particle-reservoir interaction. Computer experiments [6], [7], [34], [35], [36], [37] demonstrate a possible scenario (TZSChaos) of bottlenecked evolution for an ensemble of interacting three-level AEs. Note that this slowing down in SLR is *non-critical* and there is no necessity of fine tuning of the system. On the other hand, such the scenario of bottlenecked SLR is entirely self-organized one, so no external driving (as in SOC dynamics [41]) is required to support TZSChaos.

Spatio-temporal dynamics of RSW-ensembles is sensitive not only to presence of absorbing boundaries in our isotropic and homogeneous nonequilibrium dissipative system, but to the relationship between a linear dimension $d_L$ of the active medium and the Wiener wavelength $\lambda_W$ (by definition, $\lambda_W$ is equal to the distance between neighbouring coils of a RSW). For moderate values $d_L/\lambda_W$ (of order 10), there exist robust spiral autowaves with $|Q_T| \geq 1$. But for large values $d_L/\lambda_W \gg 10$, only the spirals with $|Q_T|=1$ usually retain their robustness, and multispiral dissipative structures consist of spirals with this minimal topological charge (note that dissipative structures with $Q_T = 0$ are not spirals, they are called pacemakers or target patterns). This is an important difference between homogeneous [6], [7] and inhomogeneous (perforated) [37] active media, because in media with "holes" there are robust spirals with $|Q_T| \leq p_H/\lambda_W$, where $p_H$ is the perimeter of a hole.

## 7. Aggregate states of RSW-ensembles – experiments and some useful metaphors

Complex dissipative structures emerged in excitable media with 2C-diffusion gradually evolve to more or less pure multi-vortex states (formed by competing RSW). Our computer experiments have shown that these collective vortex matter states have some important features of aggregate states of the usual matter. Resulting aggregate state is first of all defined by parameters of an active medium. Preliminary (and rather metaphoric) classification of the observed aggregate states of the vortex matter in our excitable media is as follows:

- Wigner crystal state (WCS) with frozen positions of all the RSW cores, similarly to a Wigner aperiodic crystal;
- Glass state (GS) with huge quantity of attractors and superslow ageing of the spatially-temporal structure;
- Evaporating liquid state (ELS) with short range ordering, quick drift of RSW cores and decreasing quantity of RSW $K_{RSW}$, down to $K_{RSW}=1$ or even to $K_{RSW}=0$ (in the last case, the entire collapse of excitations takes place);
- Nonevaporating liquid state (NLS), which is similar to ELS, except for $K_{RSW}(t) \approx \mathrm{const}$ with $K_{RSW}(t \to \infty) \gg 1$.

- Dense plasma state (DPS) with strongly distorted (from Archimedean RSW) "ionized" autowaves, which produce complex, unstable, quickly oscillating irregular patterns of vortices of both RSW and non-RSW type.

As it was found in our computer experiments, the fully developed WCS is characterized by dynamically stable coexistence of RSW, most of which has $\text{sgn}\, Q_T = \pm 1$. The typical process of dynamical stabilization of this ensemble of RSW is the simplest one as compared with formation of other states of the vortex matter. Under spatially-aperiodic starting excitation of such a medium, the pointed process usually has two qualitatively different stages – the fast (explosive) starting stage and the moderately slow stage of crystallization of an aperiodic vortex grid. This grid is formed by domains of RSW with shocking waves as domain walls. The resulting spatially-irregular vortex grid is robust against external small but finite perturbations. Time-periodically forced WCS demonstrate acoustic-like motions over the active medium.

Moreover, under spatially-periodic starting excitation of such a medium, the regular crystalline structure of the vortex matter is formed, and the phonon-like states are generated by a subsequent time-periodic external forcing (likewise to generation of coherent acoustical phonons in usual solid state crystals).

Most of GS at short time intervals looks as RSW-ensembles with frozen cores (similarly to stabilized WCS), but at large time-scales the patterns of RSW evolve in a very complex and special manner. There are many different stages during GS emergence and the quantity of attractors is extremely high. So, the complexity of evolution of the GS vortex matter is, at least at intuitive level, much higher than the complexity of evolution of the WCS vortex matter.

In contrary, evolution of ELS usually is ended by the "winner takes all" state (the single RSW is occupied all the active medium) or by the simple collapse of excitations over all the active medium (no RSW and no excited/refractory AEs are here at all). On the other hand, intermediate stages of evaporating vortex liquid are of great interest because they include processes of self-induced inversion of $\text{sgn}\, Q_{\Sigma T}$, where $Q_{\Sigma T}$ is the sum of $Q_T$ over the space of the active medium. Such a counter-intuitive phenomenon was observed by us only for excitable media which typically evolve to the ELS.

Evolution of NLS proceeds, as it was observed in our computer experiments, by the mentioned above scenario of the TZSChaos (with almost conserved $K_{\text{RSW}}$) and needs further investigations due to giant duration of transient process. Inversion of $\text{sgn}\, Q_T$ in NLS is local and fast, but $Q_{\Sigma T}$ is changed very slowly.

And, finally, DPS is not a purely vortex matter state, but vortex dynamics plays here an important role too. DPS is probably the most complex and intriguing state of excitable system, which needs separate studies.

## 8. Chimera states in RSW-ensembles – spatial coexistence of incongruous states

A special interest represents a group of phenomena of forming and decay of the so-called chimera states (coexisting spatio-temporal structures with incongruous dynamics), which were observed by us in excitable systems [6]. Chimera state does not belong to any of described above uniform aggregate states. Chimera states are known for nonlocally coupled oscillators [42], [43], i.e. systems consisting of elementary units with inherent oscillatory behavior. In contrast to the oscillatory systems, elementary units of excitable systems are monostable relaxational ones. In other words, an isolated excitable AE has the single non-oscillatory stable stationary state, so any rhythmic transitions between the levels of AE are possible only in an ensemble of interacting excitable units.

From this point of view the emergence of chimera states (e.g. the coexistence of periodically and aperiodically oscillating spatial domains [6]) is rather unexpected property of dissipative system containing only relaxational units. But our real radio-physical experiments with microwave phaser also show [24] coexistence of periodically and aperiodically oscillating spatial domains of excitable paramagnetic system $\text{Cr}^{3+}:\text{Al}_2\text{O}_3$. Consequently, both computer modeling and real radio-physical experiments demonstrate presence of chimera states in excitable nonequilibrium dissipative systems, where a special kind of competition of localized dissipative structures ensures their antagonistic but persistent coexistence.

## List of abbreviations

| | | |
|---|---|---|
| AE | – | Active Element |
| ATLM | – | Algorithm of Three-Level Model (of excitable system) |
| AV | – | Active Vicinity |
| CA | – | Cellular Automaton |
| DPS | – | Dense Plasma State |
| ELS | – | Evaporating Liquid State |
| GS | – | Glass State |
| NLS | – | Nonevaporating Liquid State |
| RSW | – | Rotating Spiral Wave |
| SLR | – | Spin-Lattice Relaxation |
| TLCA | – | Three-Level Cellular Automaton |
| TLM | – | Three-Level Model of excitable system |
| TZSChaos | – | Transient Zeno-like Spatio-temporal Chaos |
| WCS | – | Wigner Crystal State |
| ZM | – | Zykov-Mikhailov (model) |